\documentclass[useAMS,usenatbib]{mn2e}
\usepackage{graphicx,natbib,times,mathptm,bm, amssymb, bibentry, ifpdf, datetime}
\usepackage[fleqn]{amsmath}
\pdfoutput=1
\def\Halpha{{\rm H}\alpha}

\def\minnum{200}

\bibpunct{(}{)}{;}{a}{}{,}

\ifpdf
\fi

\begin{document}

\bibliographystyle{mn2e}

\title[A 3D extinction map of the Northern Galactic Plane]{A 3D extinction map of the Northern Galactic Plane based on IPHAS photometry}

\author[Sale et al.]{\parbox{\textwidth}{S.~E.~Sale$^{1}$,
J.~E.~Drew$^{2}$,
G.~Barentsen$^{2}$,
H.~J.~Farnhill$^{2}$,
R.~Raddi$^{3}$,
M.~J.~Barlow$^{4}$,
J.~Eisl\"offel$^{5}$,
J.~S.~Vink$^{6}$,
P.~Rodr\'\i guez-Gil$^{7,8}$,
N.~J.~Wright$^{2}$
}\vspace{0.4cm}
\\
\parbox{\textwidth}{$^1$ Rudolf Peierls Centre for Theoretical Physics, Keble Road, Oxford OX1 3NP, U.K.\\
$^{2}$ School of Physics, Astronomy \& Mathematics, University of Hertfordshire, College Lane, Hatfield, Hertfordshire, AL10 9AB, U.K\\
$^{3}$ Department of Physics, University of Warwick, Gibbet Hill Road, Coventry, CV4 7AL, U.K \\
$^{4}$ University College London, Department of Physics \& Astronomy, Gower Street, London WC1E 6BT, U.K.\\
$^{5}$ Th\"uringer Landessternwarte, Sternwarte 5, 07778, Tautenburg, Germany\\
$^{6}$ Armagh Observatory, College Hill, Armagh, Northern Ireland, BT61 9DG, U.K.\\
$^{7}$ Instituto de Astrof\'\i sica de Canarias, V\'\i a L\'actea, s/n, La Laguna, E-38205, Santa Cruz de Tenerife, Spain \\
$^{8}$ Departamento de Astrof\'\i sica, Universidad de La Laguna, La Laguna, E-38204, Santa Cruz de Tenerife, Spain
}
}

\date{Received .........., Accepted...........}

\maketitle

\begin{abstract}
We present a three dimensional map of extinction in the Northern Galactic Plane derived using photometry from the IPHAS survey.
The map has fine angular ($\sim~10~\arcmin$) and distance (100~pc) sampling allied to a significant depth ($\gtrsim 5$~kpc).
We construct the map using a method based on a hierarchical Bayesian model as previously described by \cite{Sale_only.2012}.
In addition to mean extinction, we also measure differential extinction, which arises from the fractal nature of the ISM, and show that it will be the dominant source of uncertainty in estimates of extinction to some arbitrary position.
The method applied also furnishes us with photometric estimates of the distance, extinction, effective temperature, surface gravity, and mass for $\sim 38$~million stars.
Both the extinction map and the catalogue of stellar parameters are made publicly available via \url{http://www.iphas.org/extinction}.
\end{abstract}

\section{Introduction}\label{sec:intro}

Dust is a significant component of the plane of the Milky Way, essential to the cycling of matter between successive generations of stars, while also posing an observational challenge through the progressive extinction of starlight.  
Both stars and interstellar dust follow their own distributions within the disk of the Galaxy -- distributions  that we wish to establish to some accuracy in order to understand how the Galaxy has been built and will evolve.
The major concentration of Galactic dust is within 100~pc or so of the Galactic plane \citep{Marshall_Robin.2006}, along with the majority of stars \citep{Juric_Ivezic.2008}.
So far we have only a poor grasp of how extinction builds up as a function of distance away from the Sun, even if there has been good progress at far IR wavelengths in determining total dust columns according to sightline \citep{Schlegel_Finkbeiner.1998, PlanckCollaboration_Abergel.2013a}.

Mapping extinction in three dimensions both removes a nuisance factor for astrophysical studies and enables the determination of the dust distribution within the Galactic disk and the physical processes shaping it.
Observations of the gas phases of the interstellar medium (ISM) and dust in continuous emission necessarily integrate over the entire column out to the `edge' of the Galaxy -- unsupported, such data cannot directly map three dimensional structures.
In significant contrast, extinction only depends on the dust column between us and the very many observed stars, which appear at a range of distances.  
Accordingly, the systematic measurement of extinction to a large number of stars can offer an unusually direct route to the three dimensional structure of the ISM.
Our location within the Galaxy means that we can study its ISM in far greater detail than that of any other galaxy.
The knowledge we gain from studying our own Galaxy can subsequently be applied to others.

The history of extinction mapping extends back to \cite{Trumpler_only.1930}.
More recently, \cite{Neckel_Klare.1980} used observations of 11,000 stars to map extinction near the Galactic Plane, whilst \cite*{Arenou_Grenon.1992} used the INCA database of 215,000 stars to constrain a 3D model of extinction.
The recent growth in survey astronomy and the resulting explosion in the volume of available data has facilitated extinction mapping, enabling significant improvements in depth, precision and accuracy.
\cite{Marshall_Robin.2006} used 2MASS observations of red giant stars to map extinction in the inner Galaxy. \cite*{Majewski_Zasowski.2011} paired 2MASS and GLIMPSE data, whilst \cite{Berry_Ivezic.2012} used SDSS in conjunction with 2MASS.
\cite{Vergely_Valette.2010} and \cite{Lallement_Vergely.2014} have applied geophysical techniques from \cite{Tarantola_Valette.1982} to map the local ISM.
\cite{Sale_only.2012} presented a method for mapping extinction using a hierarchical Bayesian model, simultaneously estimating the distance extinction relationship and the properties of the stars along the line of sight.
Subsequently \cite{Green_Schlafly.2014} have proposed a similar method.

The INT/WFC Photometric $\Halpha$ Survey of the northern Galactic Plane \citep[IPHAS; ][]{Drew_Greimel.2005} is the first comprehensive digital survey of the northern Galactic disc ($|b|\leq5^{\circ}$), covering a Galactic longitude range of $30^{\circ} \lesssim l \lesssim 215^{\circ}$.
Imaging is performed in the $r$, $i$ and $\Halpha$ bands down to $r\sim20$. 
As demonstrated by \cite{Sale_only.2012}, the IPHAS photometric system is unusually well suited to breaking the degeneracy between extinction and effective temperature: the $(r-\Halpha)$ colour is effectively a proxy for $\Halpha$ equivalent width, and thus effective temperature, and is less heavily influenced by extinction.

An initial public data release of IPHAS data obtained before the end of 2005 was previously made available \citep{Gonzalez-Solares_Walton.2008}. 
Since then, the entire survey area has been observed at least once, leaving only 8\% of the footprint still wanting survey-standard photometry.
Furthermore, a global photometric calibration of all IPHAS data has recently been completed.  
A second data release (IPHAS DR2) containing both the additional data and the new global photometric calibration of understood precision is described by \cite{Barentsen_Farnhill.2014}.
\cite{Sale_Drew.2009} demonstrated that attempting to map extinction using inconsistently-calibrated data would inevitably lead to results exhibiting erratic biases.
However, now that a global photometric calibration is in hand it is possible to utilise the IPHAS DR2 catalogue to map 3D extinction across the entire northern Galactic Plane at high angular resolution for the first time.
The aim of this paper is to apply the method of \cite{Sale_only.2012} to achieve this mapping.

The organisation of this paper is as follows.
Section~\ref{sec:method} summarises the method employed, concentrating on departures from \cite{Sale_only.2012} and details particularly relevant to the large scale implementation we perform here.
In section~\ref{sec:map} we present the 3D extinction map alongside examples of derived stellar parameters.
We then compare our extinction map to those of \cite{Marshall_Robin.2006} and \cite{Schlegel_Finkbeiner.1998}, before closing with a discussion of the map produced here and future work.

\section{Method}\label{sec:method}

\cite{Sale_only.2012} proposed a method for simultaneously mapping extinction and determining stellar parameters using a hierarchical Bayesian model.
The Markov Chain Monte Carlo (MCMC) based algorithm, called H-MEAD (hierarchically mapping extinction against distance), simultaneously estimates the distance-extinction relationship along a line of sight and the properties of the stars sampled.
The first step in this is to define an angular resolution element -- an area on the sky for which we wish to find a distance-extinction relationship.
Then, for each angular resolution element, we create a series of voxels\footnote{a voxel is conventionally a volume pixel} by dividing the conical volume projected onto each 2D angular resolution element into distance slices, each 100~pc in depth.
We then employ IPHAS photometry of stars within the angular resolution element as the means to estimate the distance-extinction relationship as well as the characteristics of the stars. 

Dust in the ISM is shaped by turbulent flows, that form fractal structures within the ISM across a range of scales from kpc to $\sim 100$~km \citep{Spangler_Gwinn.1990, Chepurnov_Lazarian.2010}.
This gives rise to differential extinction: since the dust density is likely to vary on very small scales -- even within an already small but finite angular element -- extinction to a given distance (or voxel) will take a range of values.
The treatment of \cite{Sale_only.2012} includes differential extinction in the statistical model, accounting for its influence on observations and enabling it to be measured for the first time.
This is a key advance relative to many other extinction mapping techniques, such as those of \cite{Marshall_Robin.2006} and \cite{Green_Schlafly.2014}.  

Formally, H-MEAD estimates the posterior distribution 
\begin{align}
P(\bm{x}, \bar{\bm{ A}}|\bm{\tilde{y}}) \propto P(\bm{\tilde{y}}|\bm{x})P(\bm{x}|\bar{\bm{ A}})P(\bar{\bm{ A}}), 
\label{eqn:posterior}
\end{align}
\noindent where $\bm{x}$ represents the properties of the stars within the angular resolution element, including distance, extinction, stellar effective temperature, surface gravity and metallicity. 
$\bar{\bm{ A}}$ is the distance-extinction relationship, where the extinction to a given voxel at a given distance  
is described by a lognormal distribution, parametrized by a mean extinction and a standard deviation.
As a shorthand we will refer to the standard deviation of the extinction to a given voxel as the \textit{differential extinction}.
Finally $\bm{\tilde{y}}$ represents the observations we have -- in this case photometry from the IPHAS DR2 catalogue.
Therefore, we are attempting to find the posterior probability distribution of the stellar properties and the extinction-distance relationship consistent with the IPHAS DR2 catalogue.
By repeating this analysis for many angular resolution elements we acquire a large 3D map of extinction.

The relationship between $\bm{\tilde{y}}$ and $\bm{x}$ is dependent on a set of isochrones, we employ those of \cite{Bressan_Marigo.2012}, and an extinction law, in this case the $R_V=3.1$ law of \cite{Fitzpatrick_Massa.2007}.
$P(\bm{x}|\bar{\bm{ A}})$ includes priors on the distance, age, mass and metallicity of stars.
These arise from an assumed stellar density, star formation history, initial mass function and a probabilistic dependence of metallicity on Galactocentric radius.
The priors are described in more detail in \cite{Sale_only.2012}.
We note here that they are conservative and not overly constraining (e.g. they include a constant star formation rate over the history of the disk, plausible scale-lengths and heights).
In general, the only constraint given by $P(\bar{\bm{ A}})$ is that mean extinction should increase with respect to distance -- although, as discussed in section~\ref{sec:distant}, for the most distant voxels it is necessary to impose an additional constraint.

\begin{figure}
\includegraphics[angle=270]{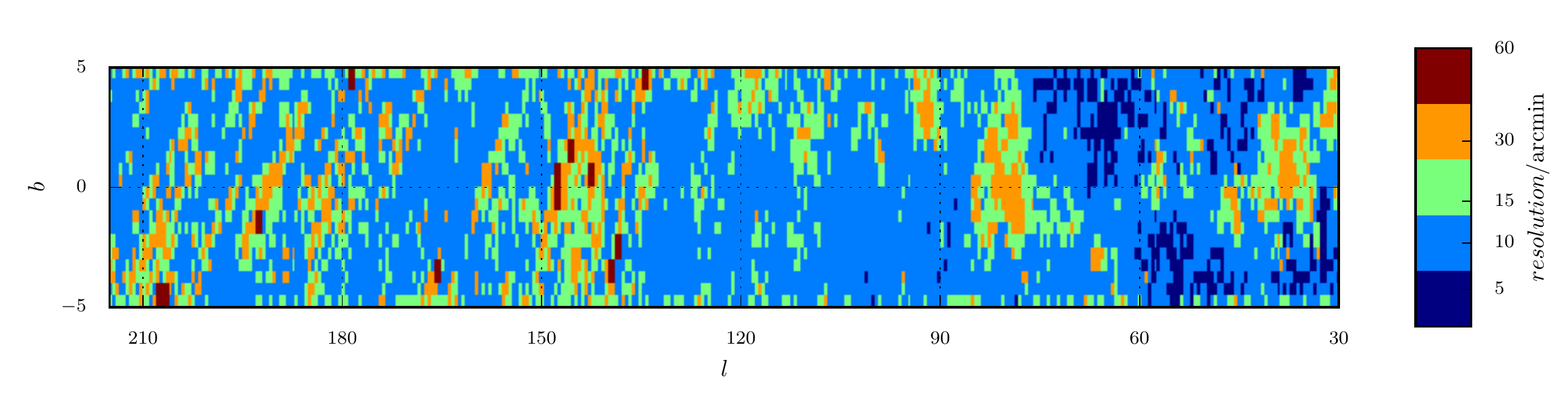}
\caption{The angular resolution of the extinction map. \label{fig:res}}
\end{figure}

\subsection{Angular resolution}

The area density of stars varies substantially across the IPHAS survey area due to variations in the three dimensional distribution of both dust and stars.
Typically the angular distribution of stars is at its densest at lower Galactic longitudes (i.e. nearer the Galactic centre) and a few degrees removed from the Galactic Plane.
In this inner Galactic region the three dimensional stellar density is high, whilst the dust, having a shorter scale height than the stars, obscures many stars near the midplane.
The observed stellar density is lowest either near the Anticentre or in regions of very strong extinction.

There is no clear and absolute standard by which we should set the angular resolution of our map.
There exists a tension between finer resolutions which capture the fractal ISM in better detail and a coarser map which will reach a somewhat greater extinction depth and is more robust against the inclusion of poor data.
This balance clearly shifts between regions of high stellar column density, where finer resolutions are more feasible, and less densely populated regions, where we have to opt for lower resolutions.
Consequently in this paper we adopt a variable resolution.
Specifically we set the resolution by ensuring that each angular resolution element contains at least $\minnum$ stars that satisfy the colour cuts we impose.
We further opt for angular resolution elements that are square, with a side of $5\arcmin$, $10\arcmin$, $15\arcmin$, $30\arcmin$ or $60\arcmin$.
By applying these criteria we arrive at a typical resolution of $10\arcmin$. 
The resultant map of angular resolution is displayed in Fig.~\ref{fig:res}.

\subsection{Treatment of distant voxels}\label{sec:distant}

In each voxel beyond the most distant star within a given angular resolution element the dust-density MCMC chain is unconstrained by the data. 
Consequently it is possible for the estimated extinction to rise sharply at such distances.
This both impedes mixing in the MCMC chain and is clearly unphysical.
Therefore, in \cite{Sale_only.2012} a prior was applied to the mean distance-extinction relationship that the dust density should follow the distribution of an axisymmetric thin disc, with a scale length of 2.5~kpc and a scale height of 125~pc, following \cite{Marshall_Robin.2006}.
This prior constrains extinction at distances beyond the most distant observed star, preventing the unwanted unconstrained growth.
However, this prior also acts to remove non-axisymmetric features from the resultant extinction map, which is clearly undesirable as such features could be genuine and include, for example, spiral arms.

There exists a logical solution to this quandary: apply the prior only to more distant voxels where there is little or no relevant data.
In order to do this, in each iteration of the MCMC algorithm, the distribution of the distances to the stars is determined.
Then the prior described above is applied {\em only} to those voxels beyond the $90^{\rm th}$ percentile of this distribution.
As the distance of the $90^{\rm th}$ percentile will change between iterations, the prior is normalised by the number of voxels affected.

\subsection{Survey selection function and data employed}\label{sec:data}

IPHAS, like any other survey of the Galaxy, obtains observations of only an incomplete sample of stars within its field of view.
Moreover, the sample obtained is not representative of all stars, as the catalogue is limited only to observations of stars falling between certain bright and faint magnitude limits.
As discussed in \cite{Sale_Drew.2009} and \cite{Sale_only.2012}, this selection effect can produce a clear bias on extinction maps: stars subject to less extinction are brighter and thus more likely to be included in the catalogue.
If not accounted for, this effect would bias extinction maps to lower values of extinction than the true extent.
\cite{Sale_only.2012} briefly discussed how this could be dealt with. 
This discussion is expanded and generalised by Sale (in prep.), following a prescription given by \cite{Loredo_only.2004}.
In broad terms, the idea is to marginalise over the unknown total number of stars enclosed within the angular resolution element: the relationship between this true number and the number of stars in the catalogue must depend on the locally-relevant distance-extinction relationship.
In particular we apply a gamma distribution prior on the total number of stars along a line of sight, assuming a local stellar density of $0.04 M_{\odot}$~pc$^{-3}$ \citep{Jahreiss_Wielen.1997}.
This is motivated by the realisation that the appearance of stars within our catalogue is essentially a Poisson process.
In this circumstance the gamma distribution, as the conjugate prior, becomes the natural choice to describe the total number of stars.
The assumed local stellar density only affects the estimated distance-extinction relationship very weakly -- it has to be changed by more than a factor of 10 to produce a discernible alteration of the inferred distance-extinction relation.

The IPHAS DR2 catalogue contains observations of stars of all stellar types.
However, as noted by \cite{Sale_Drew.2009}, current stellar atmosphere models are unable to satisfactorily describe later K and M type stars.
Consequently, as in \cite{Sale_Drew.2009}, we apply a colour-cut that removes stars with colours typical of K5V or later types and a second which removes objects demonstrating significant $\Halpha$ emission. 
This is a simple and robust approach, although we note that \cite{Green_Schlafly.2014} demonstrate an alternative method which relies on estimating the Bayesian evidence for each star.

There are also, contained in the IPHAS DR2 catalogue, observations of extended objects as well as objects which appear to be noisy artefacts.
We ignore these objects and only employ those which have been classified as being both reliable and stellar sources \citep{Barentsen_Farnhill.2014}.
The good consequence of only using data classified as reliable is that very faint ($r\gtrsim 21$), potentially problematic stars are removed from the catalogue.
The final catalogue employed here contains 38,092,694 stars -- approximately 44 per cent of those flagged as reliable in the IPHAS DR2 catalogue -- with most of the loss attributable to the colour cuts employed.

There are a number of gaps in the spatial coverage of the IPHAS DR2 catalogue, where only substandard data are currently available.
In these regions the angular resolution of our extinction map is automatically degraded so that there are no angular resolution elements that contain no stars.
As a result, the larger angular resolution elements are often those that `bridge' the gaps in the spatial coverage offered by the IPHAS DR2 catalogue.
Indeed much of the striped structure visible in Fig.~\ref{fig:res}, particularly near the Anticentre, reflects the distribution of missing fields in the catalogue.

\subsection{Parametrization of extinction}

As in \cite{Bailer-Jones_only.2011} and \cite{Sale_only.2012} we parametrize extinction using $A_0$, the monochromatic extinction at $5495\AA$, following \cite*{Cardelli_Clayton.1989}.
This is because broadband measurements of extinction or reddening, such as $A_V$, $E(B-V)$ or $A_K$, are a function of both the dust column between us and the star and the star's spectral energy distribution.
In contrast, monochromatic measures such as $A_0$ have the important characteristic that they depend only on the properties of the dust column.

\subsection{MCMC scheme}

We employ an MCMC algorithm to produce samples from the posterior given by equation~\ref{eqn:posterior}, from which we can estimate the stellar parameters and extinction-distance relationship.
\cite{Sale_only.2012} used a 'Metropolis within Gibbs' (MwG) sampler \citep{Tierney_only.1994}, which, in each iteration, first attempts to alter the distance extinction relationship, before then altering the properties for each star in turn.
In order to use MwG it is necessary to tune the proposal distributions to obtain good sampling.
In \cite{Sale_only.2012} it was possible to do this by hand since only a few data sets were studied.
However, when mapping on a large scale, it is clearly not feasible to set all the proposal distributions manually. 
Therefore we employ instead an adaptive MwG sampler \citep{Roberts_Rosenthal.2009}, adjusting the parameters in blocks: those for the distance-extinction relationship in one block and a block for each star.
As the parameters are adjusted in blocks, the desired acceptance ratios are lower than if they were adjusted one by one \citep{Roberts_Rosenthal.2001}.

In \cite{Sale_only.2012} simulated data were produced on which H-MEAD was tested.
In doing so the validity of this method was demonstrated and the relative power of the IPHAS data made clear.

\section{The extinction map}\label{sec:map}

\begin{figure}
\includegraphics[angle=270]{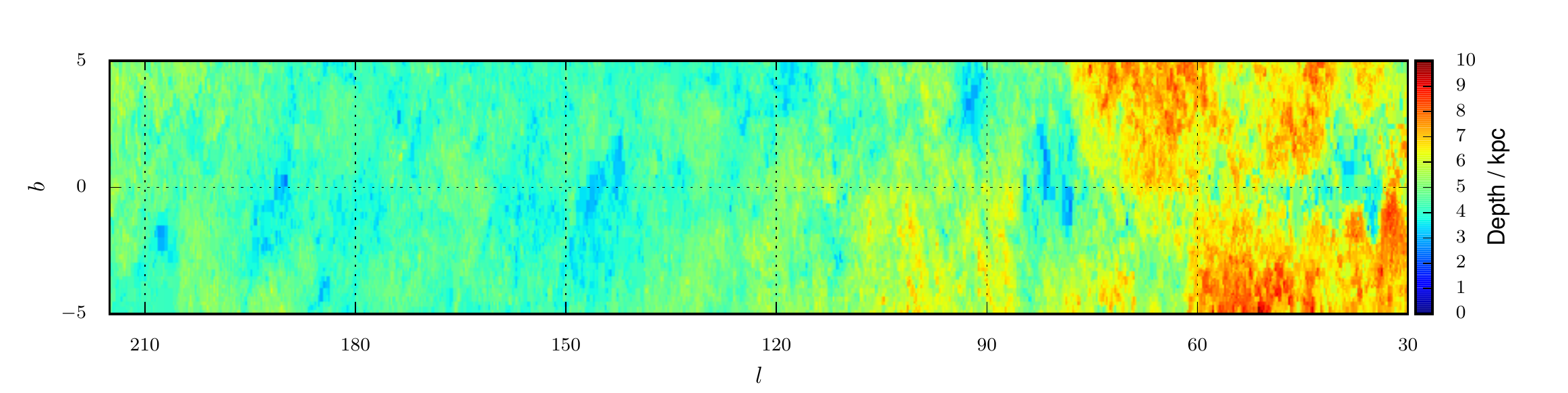}
\caption{The distance to the $90^{\rm th}$ percentile of the
  distribution of expected distances. \label{fig:depth}}
\end{figure} 

\begin{figure*}
\includegraphics[angle=270]{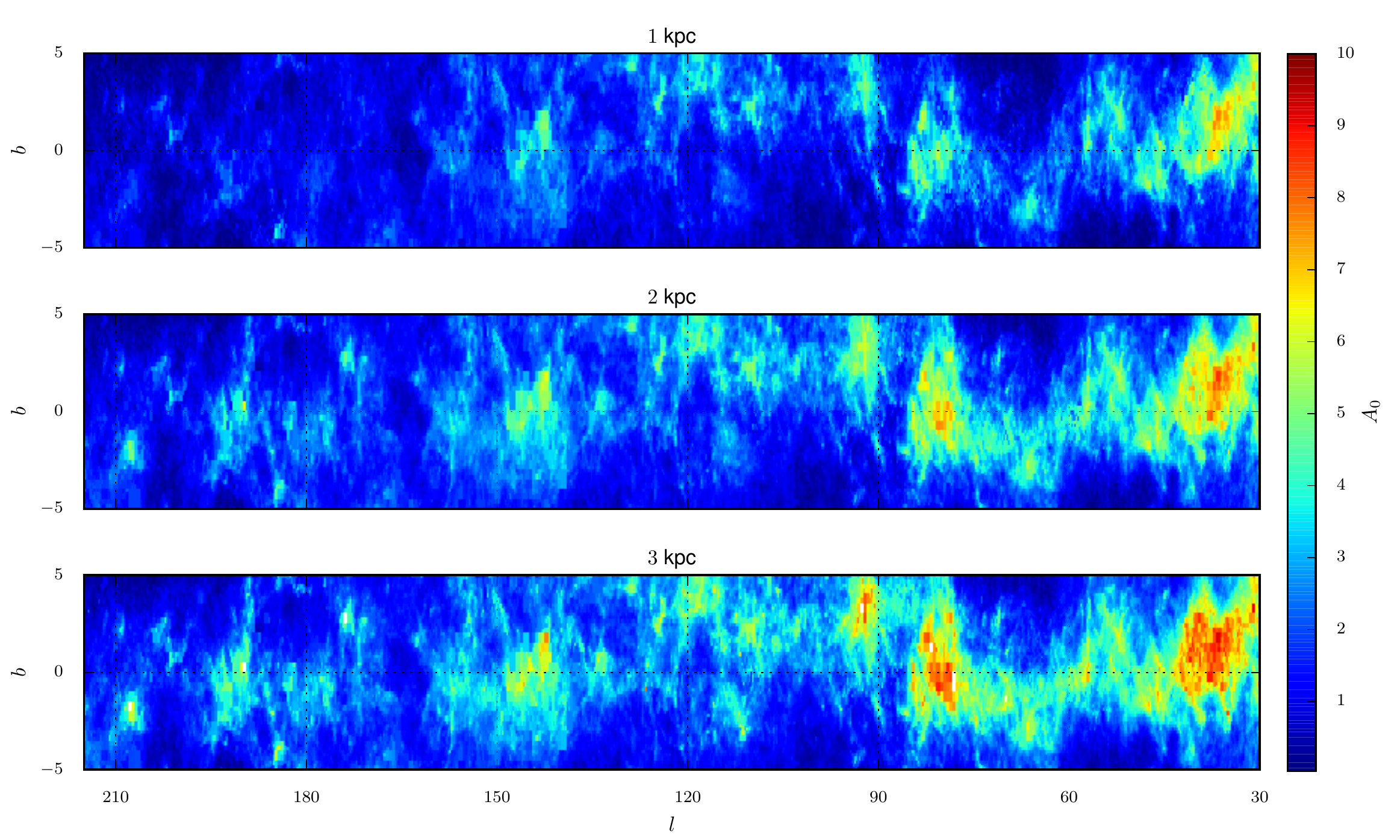}
\caption{Slices through the 3D cumulative extinction map at
  heliocentric distances of 1,~2 and 3~kpc. The extinctions plotted are from the Sun to the distance of the plane. \label{fig:map1}}
\end{figure*} 

\begin{figure*}
\includegraphics[angle=270]{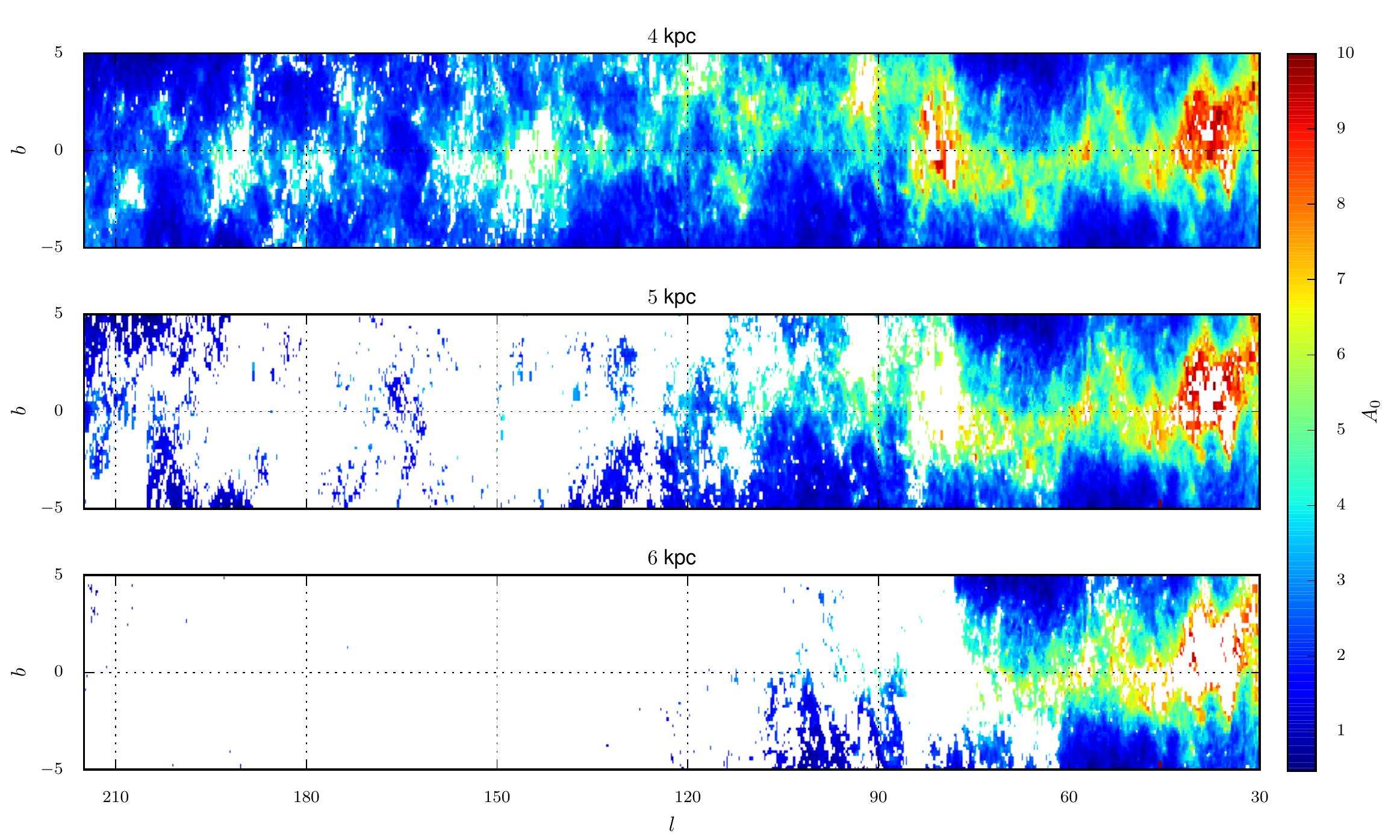}
\caption{Slices through the 3D cumulative extinction map at heliocentric distances of 4,~5 and 6~kpc.
  As in Fig.~\ref{fig:map1}, we plot extinction from the Sun to the distance of the plane. 
  The white regions are those where we are beyond the maximum reliable distance for that angular resolution element. \label{fig:map2}}
\end{figure*} 

We applied H-MEAD to the IPHAS DR2 catalogue, subject to the conditions set down in section~\ref{sec:data}, obtaining samples from the full posterior distribution given by equation~\ref{eqn:posterior}.
In what follows we will plot the mean of the posterior distribution of extinction to each voxel, whilst the quoted uncertainties on extinction are the half-width of a 68.3 per cent credible interval centred on the mean.

As discussed in section~\ref{sec:method}, we apply a prior to the most distant voxels -- those beyond almost all the stars in the IPHAS DR2 catalogue.
In a similar vein we estimate the reliable depth of our extinction map by considering the distribution of the posterior expected distances of the stars within each resolution element.
We opt to define the maximum reliable depth of the map as the $90^{\rm th}$ percentile of this distribution, in line with where the prior takes effect that prevents unphysical behaviour at the largest distances.
The map of these depths is plotted in Fig.~\ref{fig:depth} -- in the figures that follow we only plot extinction at distances within them.

Figure~\ref{fig:depth} shows the maximum reliable depth to be 5-6 kpc typically, rising to 9-10 kpc off the midplane at longitudes $\lesssim 70^{\circ}$.
Outside the Solar Circle the depth of the extinction map is more strongly limited by the extent of the Galactic stellar disc.
There are also a number of systematic features which appear as striations of the depth map and are most easily seen near the Galactic Anticentre.
These are a result of the observing strategy, which often led sequences of field acquisition and exposure at constant RA.
Hence fields observed on a night when the sky transparency and consequent faint-magnitude limit were lower can appear as a strip of relatively shallow depth.

\begin{figure*}
\includegraphics[width=\textwidth]{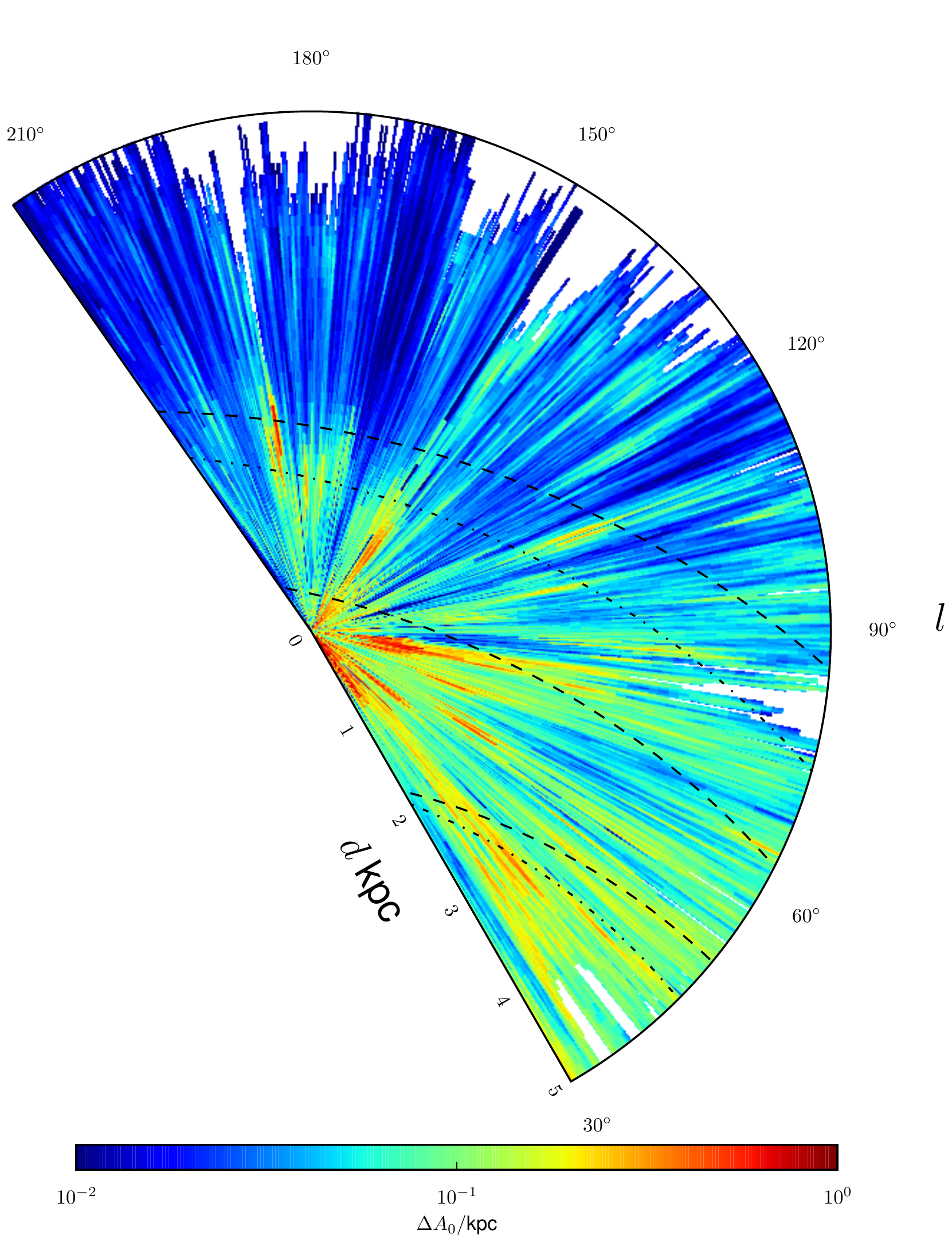}
\caption{A map of extinction at $b=0$, after convolution with a uniform kernel to reduce the resolution to $0.25^{\circ}$ in $l$ and 500~pc in distance, so as to roughly match the resolution of Figure~9 of \protect\cite{Marshall_Robin.2006}. The sun lies at the plot's origin with the Galactic centre off the bottom of the plot. The dashed lines denote the position of the Sagittarius, local and Perseus spiral arms given by \protect\cite{Reid_Menten.2014}, whilst the dot-dashed lines correspond to the Sagittarius and Perseus arms of \protect\cite{Vallee_only.2013}. Note the nonphysical `fingers of God' -- discontinuities in the azimuthal direction. \label{fig:azi_map}}
\end{figure*} 

Figures~\ref{fig:map1} and~\ref{fig:map2} show two dimensional slices of the extinction map taken at 1~kpc intervals, plotting extinction from the Sun to given distance.
These maps clearly show both the expected large scale distribution of extinction, which is generally stronger at lower latitudes and nearer the Galactic centre, and the fractal nature of the ISM.
There are a number of specific features visible in the map, for example at 1~kpc, the impact of the Aquila Rift is clearly visible at positive latitudes at $l < 40^{\circ}$, consistent with \cite{Lallement_Vergely.2014}.
Meanwhile, by 2~kpc the Cygnus Rift at $l \sim 80$ has become a clear feature, continuing to strengthen to 3~kpc.
The region of extinction corresponding to the star forming belt in the Perseus Arm, including W3 at $(l,b) \sim (134^{\circ}, 1^{\circ})$, can be seen to be a much patchier structure with a notable accumulation between 1 and 2~kpc at $l \sim 140^{\circ}$.
The Rosette Nebula at $(l,b) \sim (206^{\circ}, -2^{\circ})$ is another clear feature, appearing between 1~ and 2~kpc, as in \cite{Schlafly_Green.2014}.

\cite{Marshall_Robin.2006} previously noted that the distribution of dust is warped, an effect also visible in the map of \cite*{Schlegel_Finkbeiner.1998}.
This warp is clearly visible in Figures~\ref{fig:map1} and~\ref{fig:map2}, with the most prominent extinction in the longitude range $90^{\circ} \leq l < 150^{\circ}$ being displaced to positive latitudes.

In appendix~\ref{app:maps} we show maps of the relative uncertainty on the mean extinction.
Typically the proportional uncertainty in measured mean extinction is on the order of a few percent.
Uncertainties are generally largest within the first kiloparsec, where the number of stars in the catalogue is limited by the IPHAS bright magnitude limit of $r \sim 13$.

In Fig.~\ref{fig:azi_map} we show a slice at $b=0$ through the map of extinction {\em pseudo-density}, or the derivative of the extinction map with respect to distance.
This pseudo-density is expected to be proportional to the density of the absorbing dust.
We have degraded the resolution of this map to $0.25^{\circ}$ in $l$ and 500~pc in distance to enable a more direct comparison to be made to Figure~9 of \cite{Marshall_Robin.2006}.
This map shows a small number of over-densities which appear to be consistent with the expected position of the Perseus spiral arm as given by \cite{Vallee_only.2013} and \cite{Reid_Menten.2014}.
A slice at $b \sim 2$ would show others (see Fig.~\ref{fig:map1}).
There are also a number of over-densities, including the outer Cygnus Rift, which straddle the local arm between $l \sim 70^{\circ}$ and $l \sim 80^{\circ}$, which appears in \cite{Reid_Menten.2014} but not \cite{Vallee_only.2013}.
However, in addition there are a number of clearly nonphysical features, mostly manifested as discontinuities in the azimuthal direction (`fingers of god').
These are allied to the strong correlation in measured extinction that exists radially, but not azimuthally, between voxels. 
Their presence can make it difficult to identify real features in the distribution of dust.
However, we emphasise that these effects are less severe when considering extinction rather than pseudo-density: extinction is a cumulative measure and therefore uncertainties are proportionally smaller.
We return to this theme in the section~\ref{sec:discussion}.

\begin{figure*}
\includegraphics[angle=270]{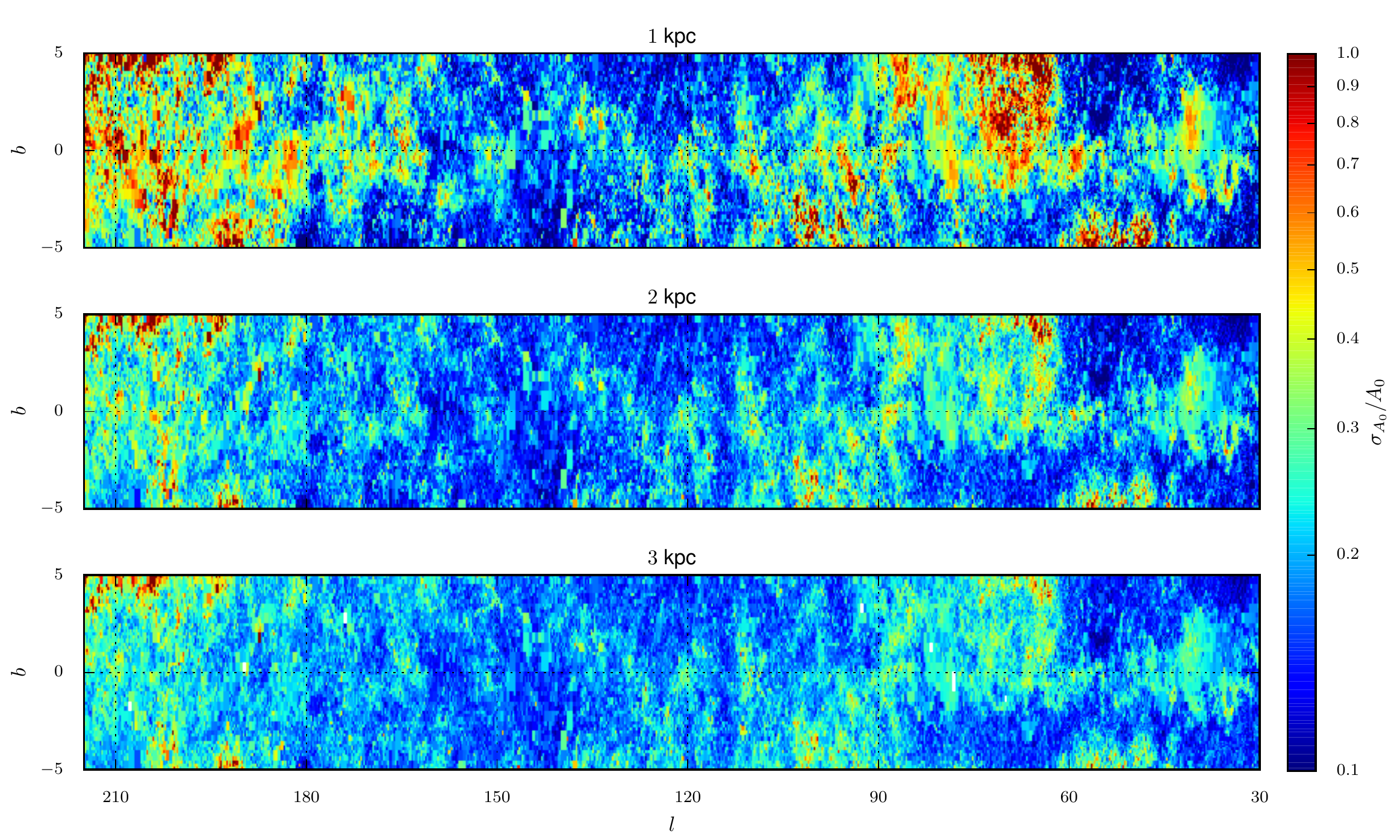}
\caption{Slices through the 3D map of the ratio of differential extinction to mean extinction at heliocentric distances of 1,~2 and 3~kpc. \label{fig:sigmaAmap1}}
\end{figure*} 

\begin{figure*}
\includegraphics[angle=270]{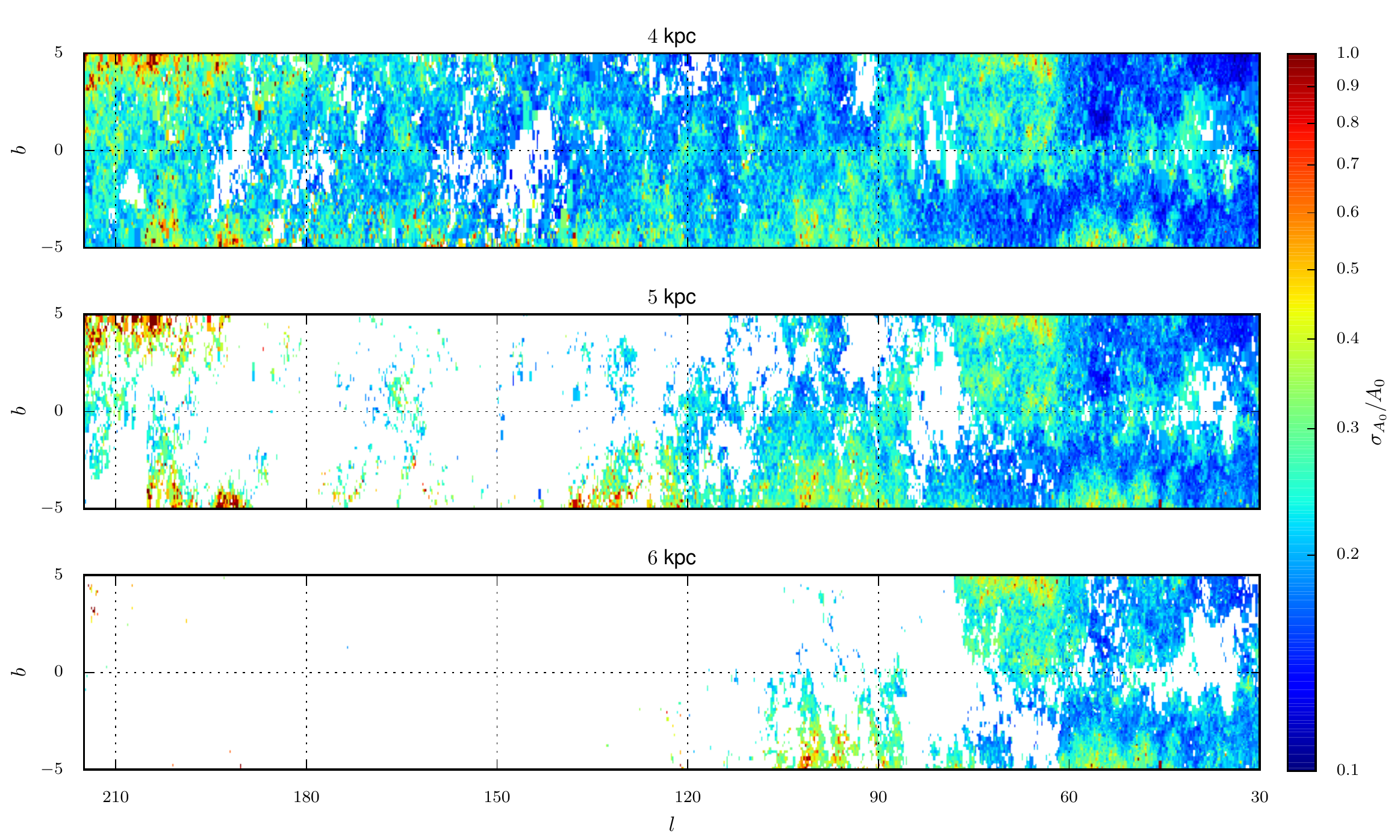}
\caption{Slices through the 3D map of the ratio of differential extinction to mean extinction at heliocentric distances 4,~5 and 6~kpc. \label{fig:sigmaAmap2}}
\end{figure*} 

Figures~\ref{fig:sigmaAmap1} and~\ref{fig:sigmaAmap2} show the ratio of differential to mean extinction.
As well as being a nuisance factor, differential extinction is also physically interesting: its extent is thought to be related to the Mach number in the ISM \citep*[e.g][]{Padoan_Jones.1997}.
As would be expected from theory \citep[e.g.][]{Lazarian_Pogosyan.2000, Fischera_Dopita.2004} and as a consequence of the outer scale of ISM turbulence being short relative to the length of our sightline, we find that the ratio of differential to mean extinction drops with increasing distance. 
We also find that the strength of differential extinction is significantly greater than the uncertainties on the mean extinction.
Therefore, although we can state the value of mean extinction within each voxel of the extinction map to high precision, we are unable to give an estimate of the extinction to a particular position within the voxel as precisely.
This has significant impact in applications such as the use of distance-extinction relationships to estimate a distance to an object of known extinction: indeed, in these applications the uncertainty will be dominated by differential extinction.

We also note that the measured differential extinctions depend on the size of the angular resolution element.
As dust density is correlated over distances up to $\sim 100$~pc, it follows that measured extinctions for two given directions will also be correlated, with the 
correlation dropping with increasing angular separation.
As a result differential extinction will be greater when the angular resolution element is larger.
Accordingly, we must stress that the measured differential extinctions are specific to the gridding of angular resolution elements employed here.

\begin{figure}
\includegraphics[]{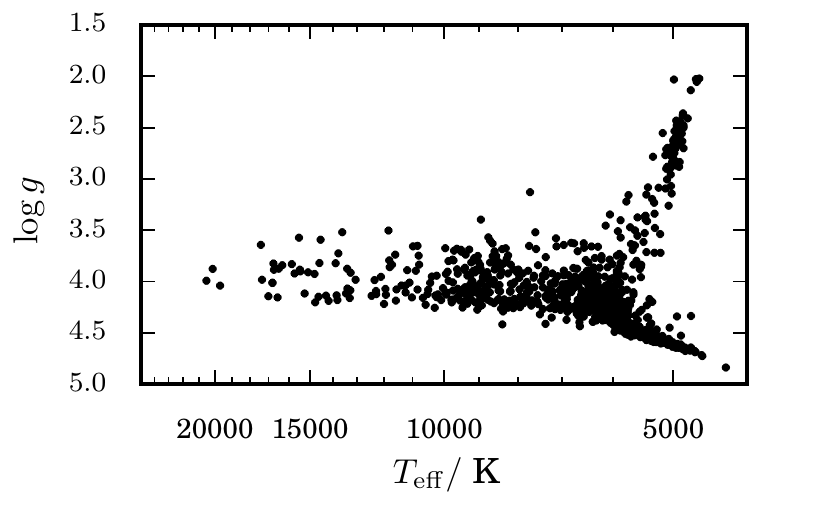}
\caption{A plot of posterior mean $T_{\rm eff}$ and $\log g$ for a field centred on $(l,b) = (80.25^{\circ}, -0.25^{\circ})$. \label{fig:logTlogg}}
\end{figure} 

In Appendix~\ref{app:data} we describe the form the extinction map takes and identify where it may be accessed.
In addition, we also provide estimated stellar parameters for the 38,092,694 stars in the IPHAS DR2 catalogue which satisfy the requirements given in section~\ref{sec:data}.
How these are made available is also presented in Appendix~\ref{app:data}.
In Fig.~\ref{fig:logTlogg} we plot the effective temperatures and surface gravities for stars in one particular resolution element.
In addition to the main sequence, the red clump and the red giant branch are both clearly visible.
This plot also demonstrates the effect of applying a colour-cut to the data: there are almost no stars with estimated effective temperatures below 4600~K, which is roughly the effective temperature of a K4V star.
The uncertainties on the measured effective temperatures and surface gravities vary, but are typically $\sim 500$~K and $\sim 0.25$ respectively for stars apparently on the main sequence.

\section{Comparison to existing extinction maps}\label{sec:analysis}

\begin{figure}
\includegraphics{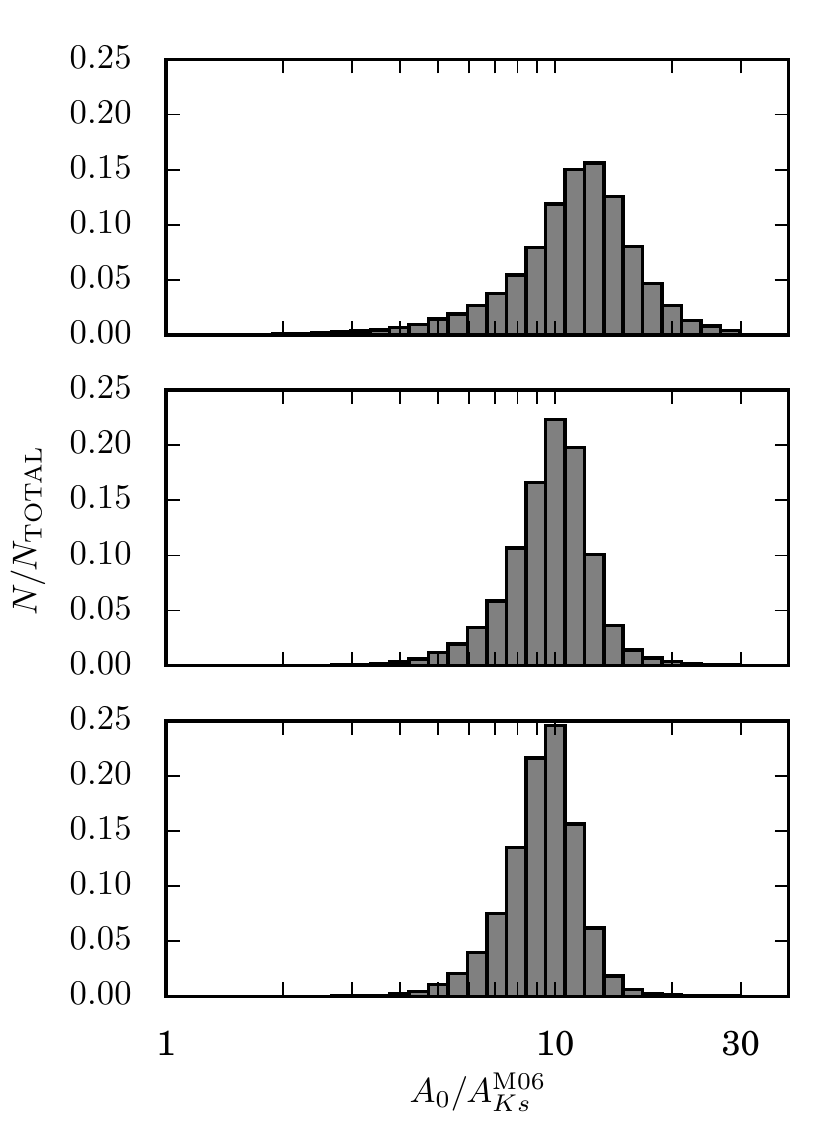}
\caption{A histogram of the ratio between the $A_0$ map produced here and the $A_{Ks}$ extinction map of \protect\cite{Marshall_Robin.2006} at heliocentric distances of, from top to bottom, 2 ,~3 and 4~kpc. \label{fig:Marshall_comp}}
\end{figure} 

We compare our extinction map to the largest and most recent 3D extinction map available -- that of \cite{Marshall_Robin.2006}.
The two maps span different regions of the sky, but do overlap in the Galactic longitude range $30^{\circ} \leq l < 100^{\circ}$.
The angular resolution of \cite{Marshall_Robin.2006} is $15\arcmin$, enabling a direct comparison to our map which operates on a similar resolution.
However their distance resolution is irregular and so in making a comparison we employ linear interpolation between the points in their map.
In Fig.~\ref{fig:Marshall_comp} we show histograms of the ratio of the two maps at three different distances.
For most directions, the closest point in the extinction map of \cite{Marshall_Robin.2006} is found beyond 1~kpc, nor do we expect great relative precision at these shorter distances in our map either.  
It is for these reasons we start the comparison at 2~kpc.

There are some significant variations in the $A_0/A_{Ks}$ ratio, which should be around 10, for an $R_V=3.1$ extinction law (\citealt{Cambresy_Beichman.2002} use $\sim$9, while \citealt{Fitzpatrick_Massa.2007}, give 11).
Some of these variations may be real and arise in regions where the shape of the extinction law is significantly non-typical.
However, it has long been accepted that Galactic reddening from $\sim 0.6$~$\mu$m up to the $K$ band is not as variable as it can be at blue and ultraviolet wavelengths \citep{Cardelli_Clayton.1989, Fitzpatrick_Massa.2007, Stead_Hoare.2009}.
If we continue to use $R_V$ to characterise reddening law, variation of this parameter through a from 2.2 up to 5.5 implies $A_0/A_{Ks}$ falling from 13.7 down to 8.5 (in the \citealt{Fitzpatrick_Massa.2007} formulation).
Hence it is likely that some of the differences are due to the different methods of analysis employed here and in \cite{Marshall_Robin.2006}.
Critically, there is no provision for differential extinction by \cite{Marshall_Robin.2006}, since their data are binned by extinction rather than by distance.
The exact impact this has on their results is unclear, although differential extinction must have left some imprint on their data given that it is a significant and pervasive feature of the ISM.
A hint that this is  an issue comes from the greater spread in $A_0/A_{Ks}$ ratio at the nearest distance plotted in Fig.~\ref{fig:Marshall_comp}, where we know that the differential extinction is the larger fraction of the mean.
It is also possible that the underpinning assumption by \cite{Marshall_Robin.2006} that red giants will follow a smooth e-folding density distribution consistent with the Besancon Galaxy Model may have helped create the tendency towards more structured (less smooth) extinction-distance relations than are typical of our results.  
In the method of \cite{Sale_only.2012} there is more freedom for the estimated parameters to be more strongly driven by the observations.

Another possibly significant difference between the two treatments is that the samples of stars used in them may be drawn from somewhat different populations, subject to differing amounts of extinction.
The contrasting wavelength ranges and colour cuts employed mean that, particularly in the most extinguished regions, the catalogue of 2MASS red giants employed by \cite{Marshall_Robin.2006} and the earlier-than-K4 IPHAS catalogue used here will be partly distinct.
The catalogue of \cite{Marshall_Robin.2006} undoubtedly offers better coverage of strong extinction at larger distances.
However, since there is some overlap between the two catalogues (both include clump giants), differences of method are certainly playing a role.
Nevertheless, Fig.~\ref{fig:Marshall_comp} does show that at 3 and 4 kpc, where both approaches are likely to be performing well, that the extinction ratio clusters satisfactorily around 10 in accord with expectation.

\begin{figure}
\includegraphics[angle=270]{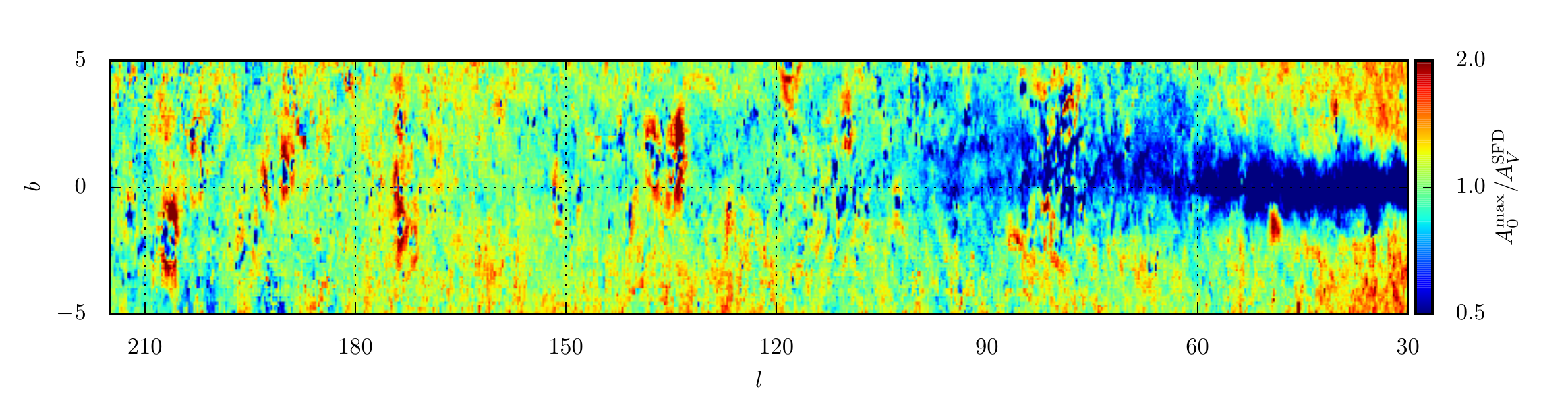}
\caption{The ratio between the estimated $A_0$ at the maximum reliable distance of our map and the asymptotic $A_{V}$ extinction map of \protect\cite{Schlegel_Finkbeiner.1998} subject to the correction of \protect\cite{Schlafly_Finkbeiner.2010}. \label{fig:Schlegel_comp}}
\end{figure}

\begin{figure}
\includegraphics{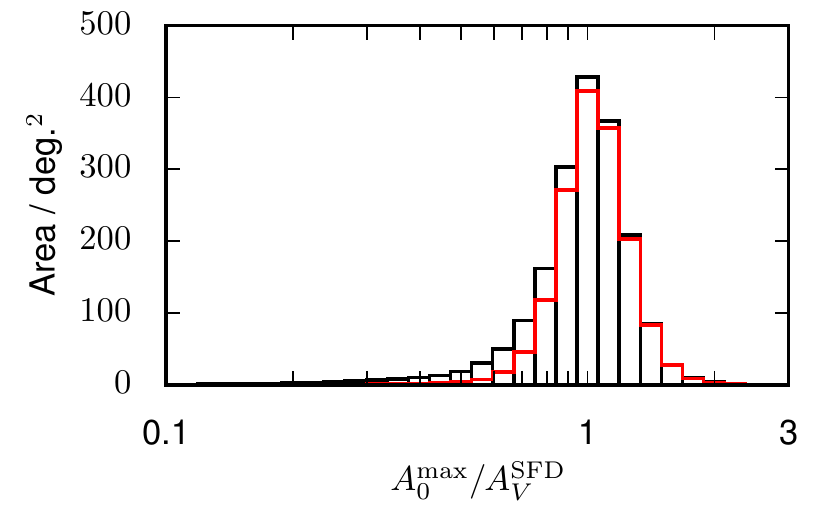}
\caption{A histogram of the ratio between the estimated $A_0$ at the maximum reliable distance of our map and the asymptotic $A_{V}$ extinction map of \protect\cite{Schlegel_Finkbeiner.1998} subject to the correction of \protect\cite{Schlafly_Finkbeiner.2010}. The black histogram includes the entire survey area, whilst the red excludes the region where $l<100^{\circ}$ and $-2^{\circ} \leq b < 2^{\circ}$.  \label{fig:Schlegel_hist}}
\end{figure}

The asymptotic extinction map of \cite{Schlegel_Finkbeiner.1998} is widely used in both Galactic and extragalactic applications, despite the fact that there are a number of known problems with it \citep*[e.g.][]{Arce_Goodman.1999, Cambresy_Jarrett.2005}.
In particular \cite{Schlegel_Finkbeiner.1998} suggest that their map {\it `should not be trusted'} at $|b| < 5^{\circ}$, which contains the entire IPHAS survey area.
Nevertheless,  we compare their map, after applying a rescaling of 0.86 following \cite{Schlafly_Finkbeiner.2010} and assuming $R_V=3.1$, to the extinction we estimate at the maximum reliable distance for each angular resolution element (Fig.~\ref{fig:Schlegel_comp}).
The depth of our map is limited by the magnitude limits of the IPHAS survey and, in the outer Galaxy and at high latitudes, by the volume density of stars.
As a result there are a number of regions of particularly high extinction where we do not penetrate the entire dust column and thus the extinction we measure at our maximum reliable depth is significantly less than the asymptotic value of \cite{Schlegel_Finkbeiner.1998}.
We refer to such regions as being `extinction limited'.
There are also a number of point sources in the \cite{Schlegel_Finkbeiner.1998} map caused by objects exhibiting infrared emission that we do not mask out.
However, across most of our map the maximum extinction we measure is approximately equal to the asymptotic extinction measured by \cite{Schlegel_Finkbeiner.1998}.
Indicating that, in general, the depth of our map extends beyond the vast majority of the Galactic dust.

Once we ignore the regions where our map is strongly extinction limited the two maps display a remarkable and reassuring agreement, particularly given that our map is entirely independent of \cite{Schlegel_Finkbeiner.1998}.
In Fig.~\ref{fig:Schlegel_hist} we show a histogram of the ratio between the two maps.
We find that the distribution of ratios strongly peaked at unity, the expected ratio given the central wavelength of the $V$ filter is very near $5495 \AA$ where we anchor our monochromatic extinctions.
When we exclude the extinction limited region  where $l<100^{\circ}$ and $-2^{\circ} \leq b < 2^{\circ}$ we find a median ratio of 1.003, with $10^{\rm th}$ and $90^{\rm th}$ percentiles of 0.81 and 1.30 respectively. 
This would appear to confirm that the correction of
\cite{Schlafly_Finkbeiner.2010}, despite being calculated on a different part of the sky, is broadly valid at low Galactic latitudes.

However, there are also a number of regions where we appear to observe stronger extinction than \cite{Schlegel_Finkbeiner.1998}.
These deviations from unity appear on spatial scales of a degree or more and exhibit clear structure.
This is larger than the sizes of our angular resolution elements, which are treated independently of each other, suggesting a physical origin for the deviations.
Indeed, several of these features correspond to regions surrounding well known H\textsc{II} regions, such as the Heart and Soul nebulae (IC 1805 and IC 1848) at $(l,b) \sim (135^{\circ}, 1^{\circ})$, the Rosette Nebula at $(l,b) \sim (206^{\circ}, -2^{\circ})$, NGC 2174 at $(l,b) \sim (190^{\circ}, 0.5^{\circ})$ and NGC 7822 at $(l,b) \sim (118^{\circ}, 5^{\circ})$.
The dust temperature in these regions will be relatively high. 
We deduce that this combines with the low angular resolution ($\sim 1^{\circ}$) of the temperature map used by \cite{Schlegel_Finkbeiner.1998} to create under-estimation of the dust column in adjacent cool regions.
This would create the evident association between H\textsc{II} regions and locations where $A_0^{\rm max} / A_V^{\rm SFD}$ is significantly greater than unity.
\cite{Schlafly_Green.2014b} observe similar features when they compare their map of extinction to 4.5~kpc to \cite{Schlegel_Finkbeiner.1998} and infer the same cause.

A further structure visible in Fig.~\ref{fig:Schlegel_comp} is the apparent warp of the extinction-limited region at $l<100^{\circ}$.
This could be the linked to warp in the Galactic dust disc that exists beyond the maximum depth of our map \citep[cf.][]{Levine_Blitz.2006b}.
In this region our map has a depth of $\sim 5$~kpc.

\section{Closing discussion}\label{sec:discussion}

We have presented a new three-dimensional extinction map of the northern Galactic Plane.
The map is produced using the IPHAS DR2 catalogue, which due to its filter set incorporating H$\alpha$ is particularly well suited to the task.
Using these data we have been able to produce a map with fine angular (typically 10 \arcmin) and distance (100~pc) resolutions.

As well as the extinction map we also produce estimates of the distance, extinction, effective temperature, surface gravity, and mass for 38,092,694 stars.
The maps and estimated stellar parameters are both being made available on line at \url{http://www.iphas.org/extinction}.

In Fig.~\ref{fig:azi_map} there are a number of azimuthal discontinuities, or `fingers of God'.  
Similar features are also visible in Fig~9 of \cite{Marshall_Robin.2006}.
These features can only exist because both methods treat different sightlines independently and are exaggerated by the fact that measured extinction is strongly correlated between voxels along a line of sight.
However, we know that no part of the ISM exists in isolation: dust density and thus extinction should be correlated between nearby regions and and it would be preferred that methods for mapping extinction would account for this correlation.

We find that the main source of uncertainty in extinction to a given position arises from differential extinction, rather than from uncertainty in the mean extinction.
This has its origins in the fractal nature of the ISM.
This is exacerbated by the way we bin the data: stars which are only a few arcseconds apart on the sky are treated in the same way as those which are further apart, but still lie in the same resolution element.
This treatment runs contrary to our physical intuition, that the extinction to closely separated stars should be more tightly correlated than that to stars more distantly spread.

\cite{Sale_Magorrian.2014} presents a method which overcomes both of these shortcomings by fitting a Gaussian random field to $\log A$, building in a physical model of interstellar turbulence.
In this way, the closest stars to a point most heavily affect the estimated value of extinction, whilst those further afield -- potentially several degrees away -- influence the estimated value of extinction to a lesser extent.
This development should lead to extinction maps without unphysical discontinuities and permit more precise estimates of extinction to individual objects.

The strength of the produced extinction map rests to a large degree on the IPHAS data employed.
The inclusion of the narrow-band $\Halpha$ filter essentially represents the cheapest conceivable spectroscopy: the combination of it with the $r$ filter allows the $\Halpha$ equivalent width to be directly estimated, offering a level of detail not usually accessed by broadband photometry.
A similar facility was available before via Str\"omgren $uvbyH\beta$ photometry.
However the clear advantages of H$\alpha$ are the penetration of longer sightlines made possible by its redder wavelength and its typically stronger signature.

The extinction map we produce is sensitive to zero-point errors in the data \citep{Sale_Drew.2009}.
The recently completed global photometric calibration of IPHAS has enabled us to produce a reliable, internally-consistent extinction map.
The comparisons we have made indicate the extinction map is astrophysically credible, which in turn underwrites the quality of the IPHAS' photometric calibration.
We can commend the use of the inferred extinction trends on spatial scales ranging from $\sim 1$~kpc up to 5-6~kpc for most sightlines (see Fig.~\ref{fig:depth}), noting that differential extinction moderates as a source of uncertainty as distance rises.
This spatial range complements that of the map of \cite{Lallement_Vergely.2014}, which provides good coverage out to 1~kpc.

The method of \cite{Sale_only.2012}, which we employ here, is not bound to any particular photometric data set.
Thus it would be possible, in the future, to combine IPHAS data with data from other surveys, such as UKIDSS-GPS \citep{Lucas_Hoare.2008} in the near infrared or UVEX \citep{Groot_Verbeek.2009}, the ongoing blue counterpart to IPHAS.
The VST/OmegaCam Photometric $\Halpha$ Survey of the Southern Galactic Plane and Bulge \citep[VPHAS+;][]{Drew_Gonzalez-Solares.2014}, which commenced in late December 2011, is the southern partner of IPHAS, observing the other half of the Galactic Plane and the Bulge.
VPHAS+ also employs the $r$, $i$ and $\Halpha$ filters, along with the $u$ and $g$ filters.
As the VPHAS+ filter set mirrors the IPHAS/UVEX filter set and the surveys share essentially the same strategy, it should be straightforward, in due course, to apply refinements to the current method to all three surveys combined in order to complete an extinction map of the entire Galactic Plane.
Looking further into the future, optical and near-infrared photometry, including $\Halpha$, collected across a range of ground-based surveys could be combined with Gaia parallaxes, as they become available, to produce much sharper, more compelling 3D maps of Galactic extinction.

\section*{Acknowledgements}

The authors would like to thank John Magorrian, Christian Knigge, Antonio Mampaso, Quentin Parker and Albert Zijlstra for constructive comments and discussions.

SES, JED and GB acknowledge support from the United Kingdom Science Technology and Facilities Council (STFC, SES ST/K00106X/1, JED and GB ST/J001333/1).  HJF acknowledges the receipt of an STFC studentship. PRG is supported by a Ram\'on y Cajal fellowship (RYC-–2010–-05762), and acknowledges support provided by the Spanish MINECO AYA2012–-38700 grant. 
The research leading to these results has received funding from the European Research Council under the European Union's Seventh Framework Programme (FP/2007-2013) / ERC Grant Agreement n. 320964 (WDTracer).

This paper makes use of data obtained as part of the INT Photometric $\Halpha$ Survey of the Northern Galactic Plane (IPHAS) carried out at the Isaac Newton Telescope (INT).
The INT is operated on the island of La Palma by the Isaac Newton Group in the Spanish Observatorio del Roque de los Muchachos of the Instituto de Astrof\'{i}sica de Canarias. 

\footnotesize{
  \bibliographystyle{mn2e}
  \bibliography{astroph_3,bibliography-2,temp}
}

\normalsize

\appendix

\section{}\label{app:data}

\begin{table*}
\begin{tabular}{c | c | c || c | c | c | c | c | c | c | c | c | c }
Name & $l / ^{\circ}$ & $b / ^{\circ}$ & $s /$~pc & $ds /$~pc & $A_0$ & $dA_0$ & $\log (T_{\rm eff} / {\rm K} )$ & $d\log(T_{\rm eff} / {\rm K} )$ & $\log g$ & $d\log g$ & $M / M_{\odot}$ & $dM / M_{\odot}$ \\
\hline
IPHAS2 J190008.19-042400.0	&	30.00006	&	-3.9367	&	4965	&	1481	&	1.60	&	0.25	&	3.782	&	0.033	&	4.41	&	0.15	&	1.02	&	0.16	\\
IPHAS2 J190009.37-042408.9	&	30.00009	&	-3.94219	&	6149	&	2222	&	2.09	&	0.49	&	3.793	&	0.059	&	4.36	&	0.17	&	1.17	&	0.26	\\
IPHAS2 J190018.64-042519.3	&	30.00011	&	-3.98536	&	4305	&	1086	&	1.45	&	0.22	&	3.767	&	0.029	&	4.43	&	0.12	&	1.03	&	0.14	\\
IPHAS2 J190021.28-042539.2	&	30.00014	&	-3.99767	&	5800	&	1691	&	1.84	&	0.33	&	3.785	&	0.039	&	4.37	&	0.14	&	1.10	&	0.19	\\
IPHAS2 J190015.86-042458.0	&	30.00015	&	-3.97239	&	3972	&	1282	&	1.36	&	0.21	&	3.763	&	0.029	&	4.38	&	0.17	&	1.03	&	0.14	\\
IPHAS2 J190020.90-042535.5	&	30.00034	&	-3.99578	&	4573	&	1650	&	1.66	&	0.27	&	3.787	&	0.035	&	4.34	&	0.19	&	1.09	&	0.19	\\
IPHAS2 J190006.36-042344.8	&	30.00038	&	-3.928	&	8521	&	1791	&	2.75	&	0.32	&	3.725	&	0.041	&	3.40	&	0.21	&	1.32	&	0.21	\\
IPHAS2 J190008.65-042402.1	&	30.00042	&	-3.93867	&	1024	&	224	&	1.10	&	0.18	&	3.722	&	0.024	&	4.55	&	0.09	&	0.85	&	0.07	\\
IPHAS2 J190010.71-042416.5	&	30.00074	&	-3.9481	&	1792	&	376	&	1.18	&	0.19	&	3.731	&	0.027	&	4.53	&	0.08	&	0.88	&	0.09	\\
IPHAS2 J190016.67-042501.1	&	30.0009	&	-3.97578	&	2771	&	1363	&	1.36	&	0.26	&	3.739	&	0.034	&	4.43	&	0.23	&	0.91	&	0.15	\\
IPHAS2 J190004.32-042326.9	& 	30.00099	&	-3.9182	&	1828	&	824	&	1.17	&	0.17	&	3.746	&	0.025	&	4.35	&	0.25	&	0.99	&	0.13	\\
\end{tabular}
\caption{An extract from the catalogue of stellar parameters, where $s$ represents the distance to the stars. The full catalogue is available online via \url{http://www.iphas.org/extinction} \label{tab:sp} }
\end{table*}

\subsection{The extinction map}

Both the 3D extinction map and the catalogue of stellar parameters are available at \url{http://www.iphas.org/extinction} and will also be made available via the CDS.

The extinction map is presented in two forms. First, we publish the posterior expectations of mean extinction and differential extinction for every voxel in the map.
These expectations are accompanied by measurements of the half-widths of 68.3 per cent credible intervals centred on the expectation for both mean extinction and differential extinction.
In addition we also make available 20 samples from the posterior distribution on the mean extinction.

For each angular resolution element the maximum reliable distance is also provided.
Measurements of mean extinction and other parameters for the voxels beyond the maximum reliable distance are published, but for reference purposes only.
These data points depend very heavily on the assumed priors and are not so heavily constrained by data in the IPHAS DR2 catalogue -- consequently, they should not be relied upon.

\subsection{Stellar parameters}

Estimated stellar parameters for the 38,092,694 stars in the studied catalogue are also made available.
In particular, we publish estimates of the distance, extinction, effective temperature, surface gravity, and mass for each of these stars.
For each parameter we provide the mean and standard deviation of the corresponding marginal posterior distribution.
The coordinates of each star are also included so as to facilitate cross-matching with the main IPHAS DR2 catalogue of \cite{Barentsen_Farnhill.2014}.
An extract from this catalogue is shown in Table~\ref{tab:sp}.

By publishing only expectations and one dimensional standard deviations we are passing on only a relatively crude depiction of the posterior probability distribution.
Consequently features such as multimodality of the posterior and covariance between parameters will not be apparent. 
Additionally, the calculation of the stellar parameters has been subject to both a set of priors and a survey selection function, both of which have heavily shaped the results obtained.
Any further analysis of these data will also be inevitably subject to these same priors and selection function.
In light of these factors, we caution the reader that great care must be taken not to over-interpret or mistake patterns which appear in this catalogue.
These data are provided in the spirit that they may be a useful resource for e.g. target selection for spectroscopic programmes or for the interested reader to gain insight into the origin of features in the extinction map.

\section{Maps of Uncertainty}\label{app:maps}

\begin{figure*}
\includegraphics[angle=270]{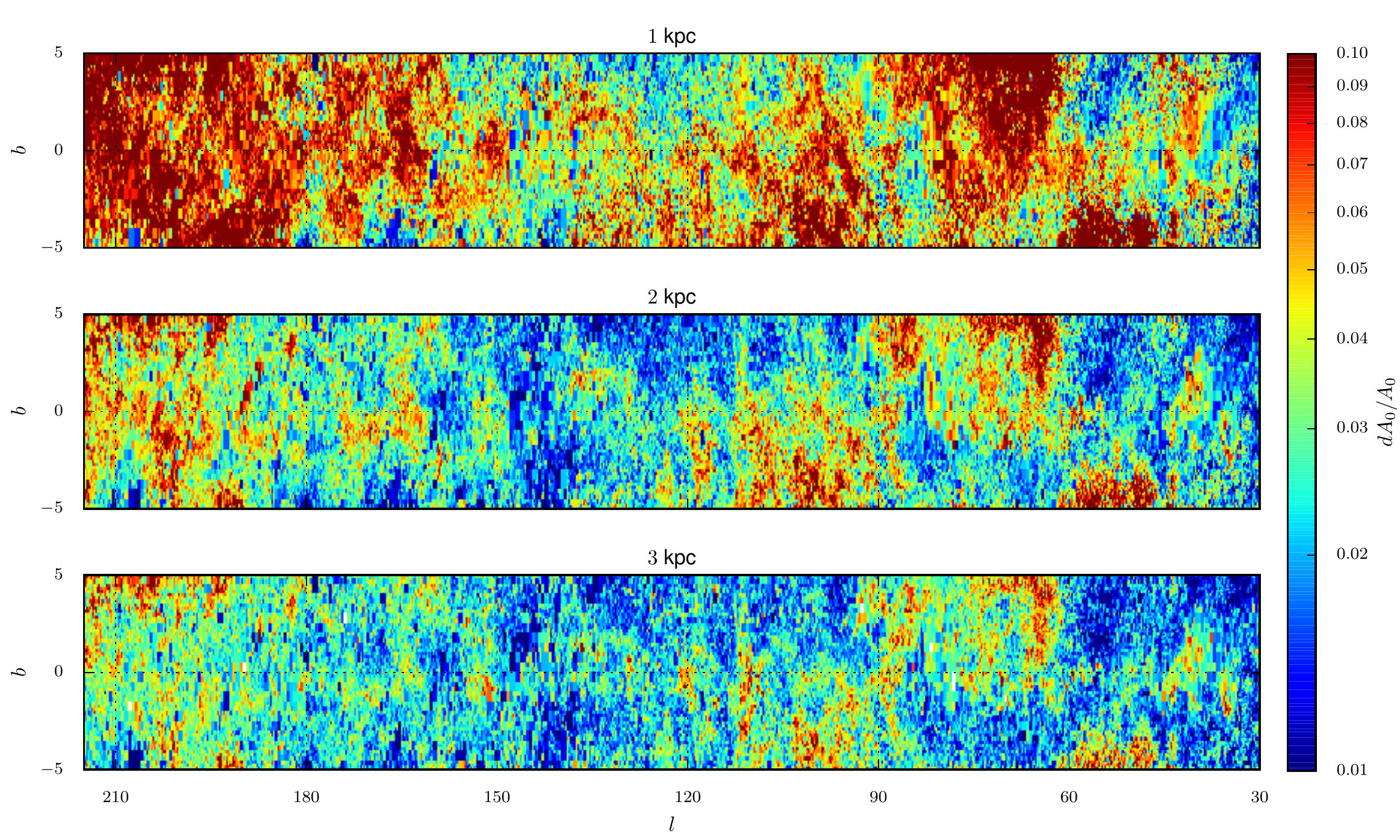}
\caption{Slices through the 3D map of relative uncertainty on mean extinction at heliocentric distances of 1,~2 and 3~kpc. \label{fig:dAmap1}}
\end{figure*} 

\begin{figure*}
\includegraphics[angle=270]{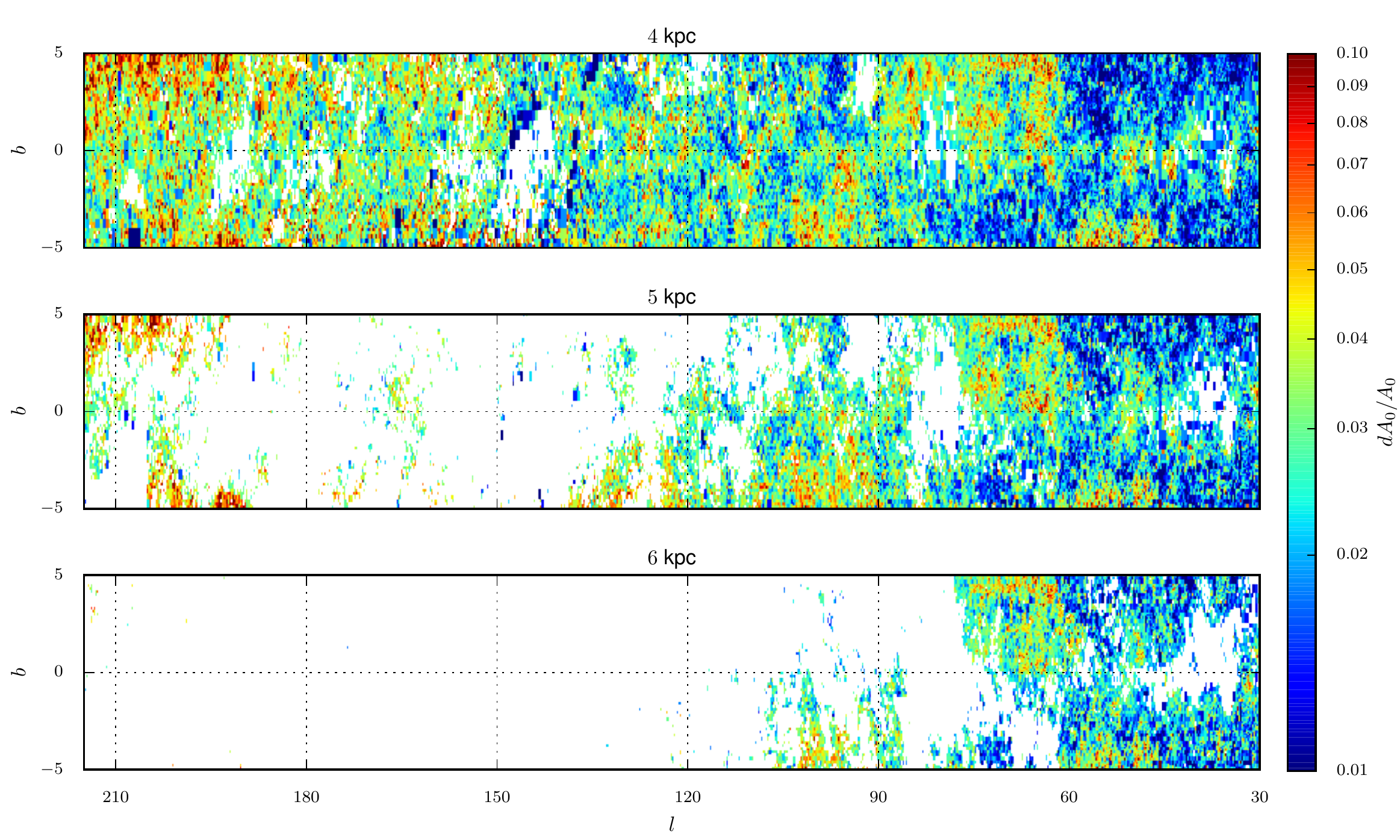}
\caption{Slices through the 3D  map of relative uncertainty on mean extinction at heliocentric distances of 4,~5 and 6~kpc. \label{fig:dAmap2}}
\end{figure*}

\end{document}